\documentclass[10pt,letterpaper]{article}
\usepackage{opex3}
\usepackage{hangcaption}
\usepackage{array}
\usepackage{indentfirst}

\newcommand{\sfft}[1]{{\rm FFT} \bigl[#1 \bigl]}
\newcommand{\sifft}[1]{{\rm FFT^{-1}} \bigl[#1 \bigl]}
\newcommand{\fft}[1]{{\rm FFT} \biggl[#1 \biggl]}
\newcommand{\ifft}[1]{{\rm FFT^{-1}} \biggl[#1 \biggl]}
\newcommand{\ffth}[2]{{\rm FFT}^{#1} \biggl[#2 \biggl]}

\begin{document}




\title{Gigapixel inline digital holographic microscopy using a consumer scanner}

\author{
Tomoyoshi Shimobaba,$^{1*}$
Hiroya Yamanashi,$^1$
Takashi Kakue,$^1$
Minoru Oikawa,$^{1}$
Naohisa Okada,$^{1}$
Yutaka Endo,$^{1}$
Ryuji Hirayama,$^{1}$
Nobuyuki Masuda,$^2$
and 
Tomoyoshi Ito$^1$
}

\address{$^1$ Graduate School of Engineering, Chiba University, 1-33 Yayoi-cho, Inage-ku, Chiba 263-8522, Japan}
\address{$^2$ Faculty of Engineering, Nagaoka University of Technology, 1603-1 Kamitomioka, Nagaoka, Niigata 940-2188, Japan}
\email{$^*$ shimobaba@faculty.chiba-u.jp}


\begin{abstract} 
We demonstrate a gigapixel inline digital holographic microscopy using a consumer scanner.
The consumer scanner can maximally scan an A4 size image (297mm $\times$ 210mm) with 4800 dpi ($\approx 5.29 \mu$m), theoretically achieving a resolution of $56,144 \times 39,698 \approx 2.22$ gigapixels.
The system using a consumer scanner has a simple structure, compared with synthetic aperture digital holography using a camera mounted on a two-dimensional moving stage.
In this demonstration, we captured an inline hologram with $23,602 \times 18,023$ pixels ($\approx$ 0.43 gigapixels).
In addition, to accelerate the reconstruction time of the gigapixel hologram and decrease the amount of memory for the reconstruction, we applied the band-limited double-step Fresnel diffraction to the reconstruction. 
\end{abstract}


\


\section{Introduction}
\noindent Digital holographic microscopy (DHM) captures a hologram with an electronic device such as a CMOS or CCD camera, and the captured hologram is reconstructed on a computer using diffraction calculation \cite{dh, kim}.

In order to increase the field-of-view, lateral and depth resolution powers of the reconstructed image, we need to capture a large hologram, for example, the amount of gigapixels achieved in recent researches,  \cite{giga1,giga2,giga3}.

There are several methods available to capture a gigapixel hologram.
In astronomy, a single CCD device with over 0.1 gigapixels has been achieved; however such a large area CCD is expensive.
Another method for acquiring a gigapixel hologram is synthetic aperture digital holography using a camera mounted on a two-dimensional moving stage (or moving a reference light).
Ref. \cite{giga3} used a green light source and a color CMOS sensor, which generally has a Bayer filter with one red, two green and one blue pixels for the acquisition of color images.
A hologram captured with the CMOS sensor and green light source lacks the pixels corresponding to the red and blue pixel due to the Bayer filter.
The absent pixels are interpolated by adjacent pixels. 

Recently, gigapixel microscopy using a consumer scanner has been proposed \cite{scanner}. 
The approach is excellent because the microscopy has a wide field-of-view, low-cost and simple structure; however, it cannot observe a sample in the depth direction. 

In this paper, we demonstrate a gigapixel inline DHM using a consumer scanner, inherently observing a sample in three-dimensions.
The consumer scanner can maximally scan an A4 size image (297mm $\times$ 210mm) with 4800 dpi ($\approx 5.29 \mu$m), theoretically achieving a resolution of $56,144 \times 39,698 \approx 2.22$ gigapixels.
The DHM system using a consumer scanner has a simple structure, compared with synthetic aperture digital holography.
In this demonstration, we captured an inline hologram with $23,602 \times 18,023$ pixels ($\approx$ 0.43 gigapixels).
In addition, to accelerate the reconstruction time of the gigapixel hologram and decrease the memory usage for the reconstruction, we applied the band-limited double-step Fresnel diffraction (BL-DSF) \cite{bldsf} to the reconstruction. 
In Section 2, we explain the gigapixel inline DHM using a consumer scanner and BL-DSF.
In Section 3, we present the results.
Section 4 concludes this work.

\section{Gigapixel inline digital holographic microscopy using a consumer scanner}

\begin{figure}[htb]
\centerline{\includegraphics[width=10.5cm]{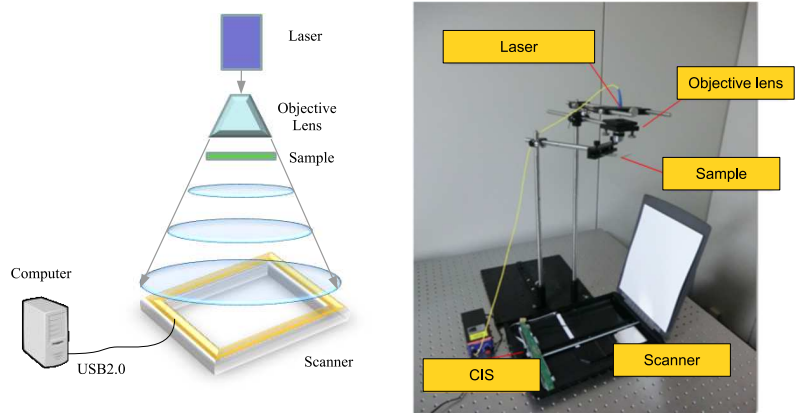}}
\caption{Gigapixel inline digital holographic microscopy using a consumer scanner. (Left) Outline of the system. (Right) Photograph of the system.}
\label{fig:system}
\end{figure}

\noindent Figures \ref{fig:system}  (a) and (b) show the outline and photograph of the gigapixel inline DHM using a consumer scanner, respectively.
Inline DHM \cite{inline1,inline2} is capable of obtaining a hologram without using beam splitters and mirrors.
Samples are placed between a light source and the hologram. 
The diffracted and undiffracted lights by the samples are regarded as the object light and the reference light, respectively, then, the interference fringe of these lights generates a hologram.
Inline DHM can control the area of the hologram and the magnification of the object light just by the location of the light source or the numerical aperture (NA) of an objective lens.
If using a holographic setup that requires beam splitters, lens and mirrors, a large aperture is needed because the scanner has a large aperture; therefore, the inline DHM is suitable for gigapixel holograms.

We used a fiber-output laser with a wavelength of 405 nm.
Although the output of the laser has a spherical wave, we used an objective lens with a magnification of $\times$10 and an NA of 0.25, to expand the angle of spread of the fiber-output laser.
We placed samples between the objective lens and the scanner.
The scanner captures an inline hologram by moving the image sensor of the scanner.    
 
\subsection{Scanner}
\begin{figure}[htb]
\centerline{\includegraphics[width=13cm]{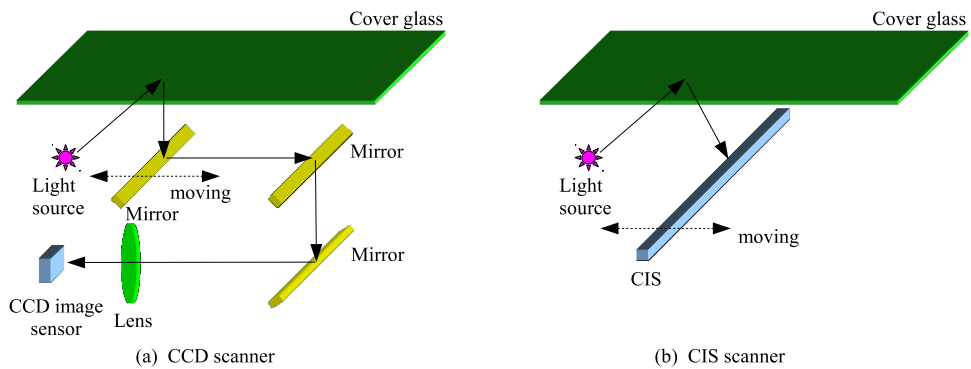}}
\caption{Two types of consumer scanners. (a) CCD scanner (b) CIS scanner.}
\label{fig:scanner}
\end{figure}

\noindent As shown in Fig.\ref{fig:scanner}, consumer scanners are mainly categorized by two types: ``CCD'' and ``Contact Image Sensors (CIS)'' scanners.
Regarding CCD scanners (Fig.\ref{fig:scanner}(a)), they have a two-dimensional CCD sensor whose size is smaller than the scan surface (cover glass); therefore, a reduction optical system composed of some mirrors and a lens is required to reduce the image on the scan surface to the CCD.
And CCD sensors generally have a color filer for scanning color images; however, the color filter disturbs the capturing of holograms \cite{giga3} .   

In contrast, CIS scanners (Fig.\ref{fig:scanner}(b)) have a simpler structure than CCD scanners because CIS sensors are one-dimensional devices whose size is the same as the scan surface; therefore, reduction of the optical system in CCD scanners is not required.  
In addition, because CIS sensors are one dimensional, the electronic circuit is simpler than CCD sensors; therefore, the CIS scanners can capture an image with 16bit / pixel maximally, while most CCD scanners capture an image only with 8bit or 12bit  / pixel maximally.
CIS sensors also do not need a color filter unlike CCS scanner because CIS sensors capture a color image by switching RGB light sources in time-divisions; therefore, CIS scanners are not adversely affected by the color filter.
Thus, we adopted a CIS scanner in this research because CIS scanners are suitable for hologram recording.

\subsection{Band-limited double step Fresnel diffraction}
\noindent The reconstruction of gigapixel holograms is very time-consuming and requires huge memory usage.
We adopted BL-DSF \cite{bldsf}, which is an effective way to obtain a reconstructed image from  gigapixel holograms.

In Fourier optics, diffraction calculations are categorized by two forms: the first is convolution-based diffraction and the second is Fourier transform-based diffraction.

Here, we show the angular spectrum method (ASM) as an example of convolution-based diffraction.
ASM is expressed as follows:
\begin{equation}
u_2(x_2, y_2)=\ifft{\fft{u_1(x_1, y_1)} \exp( - 2\pi i z \sqrt{1/\lambda^2-f_x^2-f_y^2})},
\label{eqn:conv_form}
\end{equation} 
where $\lambda$ is the wavelength of light, the operators $\sfft{\cdot}$ and $\sifft{\cdot}$ are the fast Fourier and inverse fast Fourier transform respectively,  $u_1(x_1, y_1)$ and $u_2(x_2, y_2)$ indicate a source and destination planes, $(f_x, f_y)$ is the coordinate on the frequency domain and $z$ is the propagation distance.  
The merit of convolution-based diffraction is that the sampling rates on the source and destination planes are the same; however, the demerit is the need to expand the source and destination planes by zero-padding to avoid aliasing. 
It takes large memory usage and long calculation time.

To overcome this problem, double-step Fresnel diffraction (DSF) has been proposed \cite{dsf}.
It calculates the light propagation between the source to destination planes by twice the Fourier transform-based diffraction, via a virtual plane $(x_v, y_v)$.
DSF does not need zero-padding because DSF is based on Fourier transform-based diffraction.
In addition, although most Fourier transform-based diffractions change the sampling rates on the source and destination plane, DSF does not change them.
Due to the original DSF incurring the aliasing noise under certain conditions, we proposed BL-DSF introducing the rectangular function for band-limitation to the original DSF.
BL-DSF is expressed as follows:
\begin{eqnarray}
u_2(m_2, n_2) &=& 	C_{z_2}
		\ffth{sgn(z_2)}{ 
			\exp(\frac{i \pi z (x^2_v+y^2_v)}{\lambda z_1 z_2}) 
			{\rm Rect}(\frac{x_v}{x_v^{max}}, \frac{y_v}{y_v^{max}})  \nonumber \\
&&			\ffth{sgn(z_1)}{u_1(m_1,n_1) \exp(\frac{i \pi (x^2_1+y^2_1)}{\lambda z_1} )  }
		}
\label{eqn:dsf}
\end{eqnarray} 
where $z_1$ is the propagation distance from the source plane to the virtual plane,  $z_2$ is the propagation distance from the virtual plane to the destination plane, $C_{z_2}=\exp(\frac{i \pi}{\lambda z_2}(x^2_2 + y^2_2)) $.
The operator ${\rm FFT}^{sgn(z)}$ means the forward FFT when the sign of $z$ is positive and the inverse type when it is negative.
See Ref. \cite{bldsf} how to determine the band-limiting area $(x_v^{max}, y_v^{max})$.
BL-DSF was implemented in our open-source wave optics library, CWO++ \cite{cwo}.

\section{Results}
\begin{figure}[htb]
\centerline{\includegraphics[width=13cm]{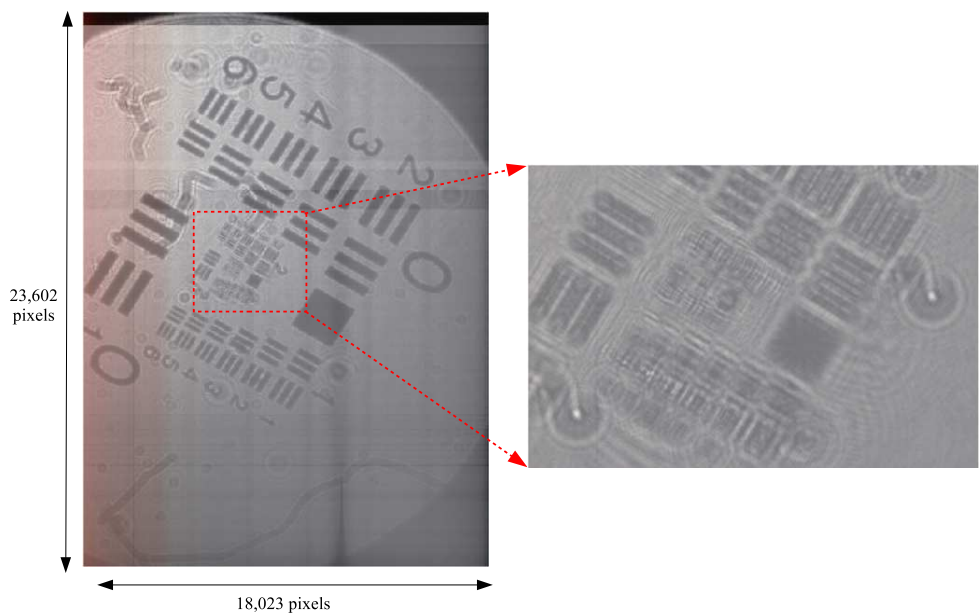}}
\caption{Hologram of USAF1951 test target with $23,602 \times 18,023$ pixels and the hologram is sampled at about 5.29 $\mu$m.}
\label{fig:holo}
\end{figure}

\noindent We show a hologram of the USAF1951 test target captured by the DHM and the reconstructed image.
Figure \ref{fig:holo} shows the hologram with $23,602 \times 18,023$ pixels (0.43Gpixels) and the hologram is sampled at about 5.29 $\mu$m. 
The hologram is captured under the condition that the distance between the scanner and the sample is 50 cm and the distance between the sample and the light source is 15 cm. 
Because the number of pixels in the hologram is much larger than the resolution of a display (1,920 $\times$ 1,080), the decimated hologram by image processing is displayed.  
The inset shows the raw hologram in a part of the hologram (in the red square). 
We can observe the interference fringe of the hologram.
Although the scanner can obtain a hologram maximally in 16bit / pixel, we were unable to recognize the difference between the reconstructed images in 8bit and 16bit / pixel.
Therefore, we used a hologram captured by 8bit / pixel in terms of the calculation time and the memory usage.

\begin{figure}[htb]
\centerline{\includegraphics[width=13cm]{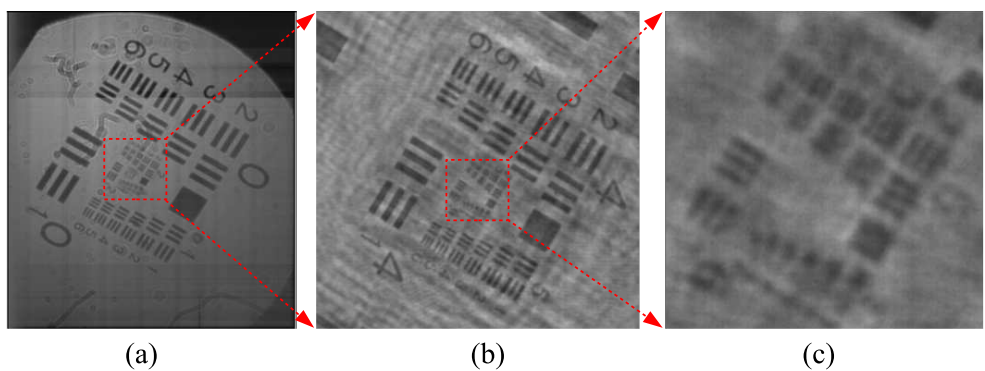}}
\caption{Reconstructed images from the hologram of USAF1951 with $23,602 \times 18,023$ pixels. (movie)}
\label{fig:reconst1}
\end{figure}

The reconstructed images are shown in Figure \ref{fig:reconst1}.
The observational area of Fig.\ref{fig:reconst1} (a) is about $22 \times 29$ mm$^2$.
Figure \ref{fig:reconst1} (b) shows the details of the red square in Fig.\ref{fig:reconst1} (a).
Figure \ref{fig:reconst1} (c) shows the red square in Fig.\ref{fig:reconst1} (b) in greater detail.
We can observe a reconstructed image in the resolution power of about $8.8 \mu$m.
The movie shows the observation of the reconstructed image as magnifying the image.

In a large hologram, the calculation time and memory usage for the reconstruction are important issues.
We compared the calculation time and memory usage in the reconstruction of the hologram (Fig.\ref{fig:holo}) using ASM and BL-DSF. 
In ASM, the calculation time and the memory usage were about 355 seconds and 12.6Gbytes, while the calculation time and the memory usage of BL-DSF were about 177 seconds and 3.2Gbytes, respectively.
The peak signal-to-noise ratio (PSNR) between the reconstructed images of ASM and BL-DSF is over 30 dB. 

\begin{figure}[htb]
\centerline{\includegraphics[width=13cm]{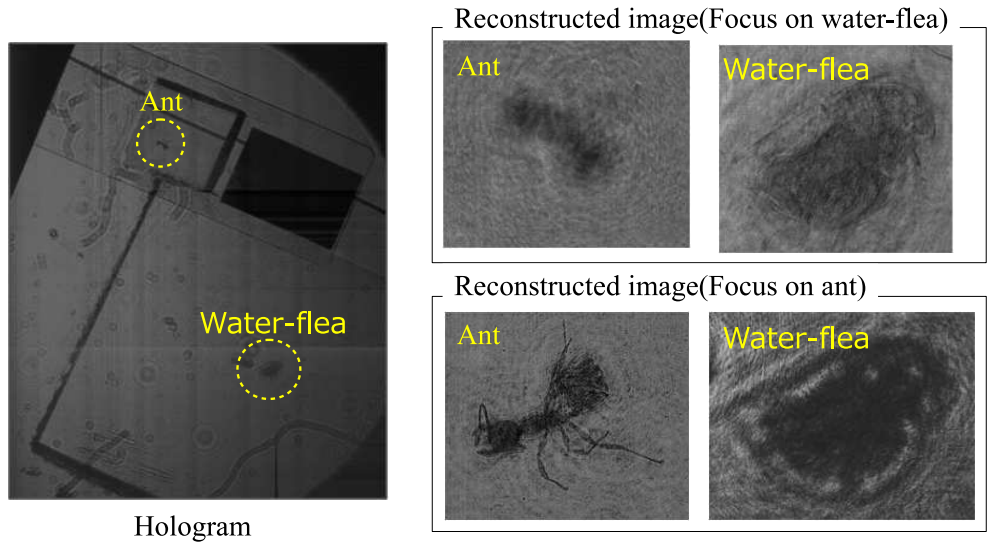}}
\caption{Hologram that, records an ant and water-flea placed at 30 cm and 50 cm from the scanner, and the reconstructed images.}
\label{fig:reconst2}
\end{figure}

Figure \ref{fig:reconst2} shows a hologram that, records an ant and water-flea placed at 30 cm and 50 cm from the scanner, respectively, and the reconstructed images.
The number of pixels of the hologram is $23,602 \times 18,023$.
When focusing on the water-flea, the ant is unfocused. 
While, when focusing on the ant,  the water-flea is unfocused.

\section{Conclusion}
\noindent We demonstrated gigapixel inline digital holographic microscopy using a consumer scanner.
In the demonstration, we showed holograms with $23,602 \times 18,023$ pixels ($\approx$ 0.43 gigapixels) captured by a consumer scanner and the reconstruction images.
We believe that the scanner is a promising imaging device for holography.
In the reconstruction of a large hologram, we showed the effectiveness of BL-DSF in terms of the calculation time and the memory usage, compared with the convolution-based diffraction.
The consumer scanner can maximally scan an A4 size image (297mm $\times$ 210mm) with 4800 dpi ($\approx 5.29 \mu$m), theoretically achieving  the  resolution of $56,144 \times 39698 \approx 2.22$ gigapixels.
We are attempting to obtain a larger hologram with our system.

\section*{Acknowledgments}
This work is supported by Japan Society for the Promotion of Science (JSPS) KAKENHI (Grant-in-Aid for Scientific Research (C) 25330125) 2013, and KAKENHI (Grant-in-Aid for Scientific Research (A) 25240015) 2013.
\end{document}